\documentclass[aps,prd,twocolumn,groupedaddress,amssymb,eqsecnum,showpacs,nofootinbib]{revtex4}
\usepackage{graphicx}
\usepackage{mathptmx}
\usepackage{bm}
\usepackage{times}

\begin{document}

\title{The influence of inhomogeneities on the large-scale expansion of the universe}

\author{Hael Collins}
\email{hael@nbi.dk}
\affiliation{%
The Niels Bohr International Academy and The Discovery Center\\
The Niels Bohr Institute, Copenhagen, Denmark}
\date{\today}

\begin{abstract}
The evolution of an inhomogeneous universe composed entirely of matter is followed from an early, nearly uniform state until the time when the inhomogeneities have begun to grow large.  The particular distribution of matter studied in this article is chosen to have a periodic variation in only one of the directions, which is simple enough to allow the behavior of the metric to be solved analytically, well beyond a linear approximation based on the initial smallness of the fluctuation.

This example provides an illustration of a universe where the inhomogeneities can affect its average expansion rate; and its simplicity allows a condition to be derived that tells when their presence should begin to become important.  Since the averages of the non-uniform parts of the metric and the matter density grow faster than their uniform parts, the average expansion rate accelerates with the advent of the era governed by the inhomogeneities.
\end{abstract}

\maketitle

\section{Introduction}
\label{intro}

One very apparent feature of the universe is its nonuniformity, which is evident even over quite large distances.  At the scale of a few hundred megaparsecs, the universe shows an intricate web of filaments and walls composed of densely clustered galaxies which are surrounded by comparatively empty voids.   These voids appear to be the largest coherent structures in the universe.  Their size, though vast enough in itself, is still but a small fraction of the observable universe today.  It is therefore often thought that inhomogeneous structures such as these voids should have little influence on the overall expansion of the universe.

One reason to question this assumption, or at least to wish to test it through specific instances, is that the equations of general relativity governing the mutual influence of matter on the shape of space-time are highly complicated and inherently nonlinear.  Although the universe has a few naturally available small parameters---the initial smallness, for example, of the fluctuations in the density of the matter, or the ratio between the typical size of a void and the size of the observable universe---it is also a dynamic and an evolving system.  The uniform part, thought to determine the overall expansion of the universe, and the spatially varying part, which describes the formation of structures such as the voids and the clusters of galaxies, grow at very different rates.  Even if the average effect of the inhomogeneities in the universe is extremely tiny at an early time, this smallness can be undone once the difference between how these two parts evolve is given enough time to show itself.

This article will demonstrate how the growth of the inhomogeneities in a universe containing only matter can affect its average expansion rate.  The example that will be used to illustrate this behavior will not be anything quite as complicated as the irregular system of walls and voids that occurs in the actual universe, but will instead consist of a universe composed of matter that has a single periodic fluctuation with a fixed comoving wavelength.  Also, the fluctuation will be chosen to depend on only one of the spatial directions.  With this simplicity, the growth of the fluctuation in this example can be solved analytically, at least until the time when the distribution of the matter becomes markedly inhomogeneous.  Beyond this point, the analytic solution becomes a poor approximation to the true behavior.  But despite its simplicity, this example is meant to capture several properties of the actual universe.  It begins in a nearly uniform state and evolves into a regular pattern of thin walls and wide voids as the matter is drawn toward the initially slightly denser regions and away from the less dense places between.

In this simplified version of the universe, it is possible to solve for the detailed time-dependence of the inhomogeneous part well beyond a linear approximation, and its periodicity means that it is straightforward to define its spatially averaged behavior.  An important result of this analysis is that it provides a way to characterize when the expansion of the universe begins to depart strongly from the behavior expected of one composed of uniformly distributed matter.  The condition that separates these phases can be stated in terms of two dimensionless parameters.  If the initial amplitude of the spatial variations in the density of the matter, as a fraction of its average density, is called $\delta$, and if the size of the observable universe measured in units of the wavelength of the fluctuation is called $n$, then the advent of the stage when the inhomogeneities strongly influence the overall expansion rate occurs when $n^2\delta\sim 1$.  The average expansion rate during the latter stage is faster than the earlier one, so as this transition occurs between them, the average expansion rate of the universe begins to speed up.  Using the measured size of the initial fluctuations in the density of the matter, if this transition were to have occurred approximately four billion years ago, the corresponding wavelength of the fluctuations today would be of the same size as the voids in the universe.  Whether this acceleration in the average expansion rate actually translates into a luminosity distance function that resembles that of the observed universe will be treated in later work, once the example here has been generalized slightly to be made to resemble more closely the observed universe.

\section{A single fluctuation}
\label{fluctuation}

From what has been observed of the universe so far, it appears to have begun in a largely homogeneous state that evolved into a truly inhomogeneous one only long after all the other known substances within it had been so diluted, in comparison with the density of the matter, that they had only a negligible role in its later expansion.  It is therefore a good approximation to neglect from the start the radiation---and any other possible ingredients---and to treat the universe as though it were composed entirely of nonrelativistic matter, which is in the form of a pressureless idealized gas.  This gas will not be perfectly uniform throughout space, but will have some spatially dependent fluctuations even at quite early times.  To simplify the analysis a little, this article will consider a case where the inhomogeneities exist in only one of the spatial directions.  Choosing a set of coordinates that is consistent with these conditions, this space-time can be described by the metric,
\begin{eqnarray}
ds^2 
&\!\!\!\!=\!\!\!\!& 
g_{\mu\nu}(t,x)\, dx^\mu dx^\nu 
\nonumber \\
&\!\!\!\!=\!\!\!\!& 
c^2(t,x)\, dt^2 - b^2(t,x)\, dx^2 - a^2(t,x)\, \bigl[ dy^2 + dz^2 \bigr] . 
\qquad
\label{metric}
\end{eqnarray}

The important property of the matter for its effect on the space-time is its density, $\rho(t,x)$.  The energy and the momentum associated with this pressureless gas is 
\begin{equation}
T_{\mu\nu}(t,x)\, dx^\mu dx^\nu = c^2(t,x)\, \rho(t,x)\, dt^2 , 
\label{matterEMT}
\end{equation}
and their conservation corresponds to the condition $\nabla^\lambda T_{\lambda\mu} = 0$.  This equation imposes two conditions on the time-evolution of the density and on the spatial dependence of the time component of the metric,
\begin{equation}
{\dot\rho\over\rho} = - 2 {\dot a\over a} - {\dot b\over b} 
\qquad
\hbox{and}
\qquad
{c'\over c} \rho = 0 .
\label{conserve}
\end{equation}
The integrated form of the first equation becomes the statement that the only time-evolution in the matter's density is that produced by its dilution through the expansion of space,
\begin{equation}
\rho(t,x) = {\rho_0(x)\over a^2(t,x)b(t,x)} .
\label{rhopartsoln}
\end{equation}
The universe is never completely empty, $\rho\not=0$, so the second of the conservation conditions indicates that the function $c(t,x)\to c(t)$ is independent of where it is being evaluated.  It can then be dispensed with altogether by redefining the time-coordinate to absorb this function, $c(t)\, dt \to dt$, leaving just
\begin{equation}
ds^2 = dt^2 - b^2(t,x)\, dx^2 - a^2(t,x)\, \bigl[ dy^2 + dz^2 \bigr] . 
\label{cmetric}
\end{equation}

How the geometry of the space-time responds to the presence of the matter is described by Einstein's equation,
\begin{equation}
R_{\mu\nu} - {\textstyle{1\over 2}} g_{\mu\nu} R = 8\pi G\, T_{\mu\nu} . 
\label{einstein}
\end{equation}
The components of this equation translate into three separate conditions on the evolution and the spatial dependence of the metric functions, $a(t,x)$ and $b(t,x)$, 
\begin{eqnarray}
{\dot a^2\over a^2} + 2 {\dot a\over a} {\dot b\over b}
- {1\over b^2} \biggl\{ 2 {a^{\prime\prime}\over a} 
+ {a^{\prime 2}\over a^2} - 2 {a'\over a} {b'\over b} \biggr\} 
&\!\!\!\!=\!\!\!\!& 
8\pi G\, {\rho_0\over a^2 b} 
\nonumber \\
{\dot a'\over a} - {a'\over a} {\dot b\over b} 
&\!\!\!\!=\!\!\!\!& 0
\nonumber \\
2 {\ddot a\over a} + {\dot a^2\over a^2} - {1\over b^2} {a^{\prime 2}\over a^2} 
&\!\!\!\!=\!\!\!\!& 
0 . 
\label{einsteinNoc}
\end{eqnarray}
Now to go further, it is necessary to choose a particular form for the comoving density, $\rho_0(x)$, and the early behavior of the metric.

The approach to be followed here is complementary to the standard treatment of the growth of the universe from a highly uniform state into a very inhomogeneous one.  Initially, the spatially dependent fluctuation in the metric and in the matter's density are both extremely small in comparison with the uniform parts of each.  In this state, the universe can be approximately described as a small inhomogeneous piece that evolves in a homogeneous background, which is often written in a standard form,
\begin{equation}
ds^2 = dt^2 - r^2(t)\, \bigl[ dx^2 + dy^2 + dz^2 \bigr] . 
\label{rtdef}
\end{equation}
It is then assumed that the growth of the inhomogeneous component---as the matter begins to accumulate into regions of higher density leaving emptier voids among them---while important at scales comparable to the size of the fluctuations, has a negligible effect on the expansion rate of the universe on much larger scales.

Yet the actual behavior might be quite different.  The relativistic dynamics governing both the overall expansion and the collapse of the matter are highly nonlinear and include parts evolving at different rates.  So while the universe has a scale which is small, the initial smallness of the fluctuations, there are other time-dependent factors, such as the relative growth rates of the homogeneous and inhomogeneous parts of the metric.  After a sufficient time, the latter can overwhelm the former.

So rather than following the growth of a full spectrum of fluctuations in a fixed background, this article will instead follow the evolution of a fluctuation with a single wavelength without separating it from the `background.'  This example will show directly how and when the growth of the inhomogeneities alters the expansion of the universe at comparatively large scales.

With this picture in mind, consider the evolution of what would initially appear to be a sinusoidal fluctuation in the density of the matter,
\begin{equation}
\rho_0(x) = \bar\rho_0 \bigl[ 1 + \epsilon\cos(2\pi kx) \bigr] , 
\label{rho0def}
\end{equation}
which would be a single Fourier mode of a more general spectrum of fluctuations.  The average comoving density of the matter is $\bar\rho_0$.  This fluctuation is meant to evolve into some of the features of the large-scale structure of the universe today.  However, the universe should have been largely uniform over all scales when, for example, the physical wavelength of the fluctuation was just equal to the size of the Hubble horizon.  So, in the beginning, the amplitude $\epsilon$ provides a good parameter for describing the metric perturbatively; to a first approximation the components of the metric do not depend on the spatial coordinates at all, and are moreover equal to each other, 
\begin{eqnarray}
a(t,x) &\!\!\!\!=\!\!\!\!& \bar a_0 t^{2/3} + {\cal O}(\epsilon)
\nonumber \\
b(t,x) &\!\!\!\!=\!\!\!\!& \bar a_0 t^{2/3} + {\cal O}(\epsilon) , 
\label{metrictozero}
\end{eqnarray}
where $\bar a_0^3 = 6\pi G\bar\rho_0$.  By the Hubble horizon, what is meant is the horizon associated with this homogeneous part, defined by 
\begin{equation}
d_H(t) = {a\over\dot a}\biggr|_{\epsilon=0} = {3\over 2} t .
\label{horizondef}
\end{equation}
Although this scale will also be later used as a yardstick, once the inhomogeneities have grown sufficiently large it becomes increasingly less meaningful as a real, physically relevant scale.

Having this small parameter available permits the metric to be calculated analytically.  Up to corrections proportional to $\epsilon^5$, the solution to Einstein's equation is 
\begin{widetext}
\begin{eqnarray}
a(t,x) &\!\!\!\!=\!\!\!\!& 
\bar a_0 t^{2/3} \biggl[ 1 + {1\over 3} \epsilon \cos(2\pi kx) \biggr] 
+ {1\over 20} {(2\pi k)^2\over \bar a_0} t^{4/3} \epsilon^2 
\biggl[ 1 - \epsilon\cos(2\pi kx) + {4\over 3} \epsilon^2 \cos^2(2\pi kx)  
\biggr] \sin^2(2\pi kx)  
\nonumber \\
&& 
- {3\over 2800} {(2\pi k)^4\over \bar a_0^3} t^2 \epsilon^4 \sin^4(2\pi kx)  
+ {\cal O}(\epsilon^5) 
\nonumber \\
b(t,x) &\!\!\!\!=\!\!\!\!& 
\bar a_0 t^{2/3} \biggl[ 1 + {1\over 3} \epsilon \cos(2\pi kx) 
- {1\over 3} \epsilon^2 \cos^2(2\pi kx) + {5\over 27} \epsilon^3 \cos^3(2\pi kx) 
- {7\over 81} \epsilon^4 \cos^4(2\pi kx) 
\biggr] 
\nonumber \\
&& 
- {1\over 20} {(2\pi k)^2\over \bar a_0} t^{4/3} \Bigl[ 
6 \epsilon \cos(2\pi kx) 
+ \epsilon^2 \bigl[ 3\sin^2(2\pi kx) - 4\cos^2(2\pi kx) \bigr] 
\nonumber \\
&&\qquad\qquad\qquad 
- \epsilon^3 \bigl[ 7\sin^2(2\pi kx) - 4\cos^2(2\pi kx) \bigr] \cos(2\pi kx) 
+ \epsilon^4 \bigl[ 11\sin^2(2\pi kx) - 4\cos^2(2\pi kx) \bigr] \cos^2(2\pi kx) 
\Bigr] 
\nonumber \\
&& 
+ {3\over 2800} {(2\pi k)^4\over\bar a_0^3} t^2 \epsilon^3 
\Bigl[ 12 \cos(2\pi kx) 
+ \epsilon \bigl[ 7\sin^2(2\pi kx) - 24\cos^2(2\pi kx) \bigr]
\Bigr] \sin^2(2\pi kx) 
+ {\cal O}(\epsilon^5) , 
\label{abtothird}
\end{eqnarray}
\end{widetext}
where the terms with the same time dependence have been grouped together.  The inhomogeneous terms are obviously growing at a faster rate, $t^{4/3}, t^2, \ldots$ , than the homogeneous one, $t^{2/3}$.  So while they start quite small, being suppressed by various powers of $\epsilon$, with enough time they eventually become the more important part for the expansion of the universe.  

To gain a clearer insight into the precise time at which the nonlinear, inhomogeneous regime begins, the time coordinate can be replaced with a more natural unit.  The comoving wavelength of the fluctuation provides one natural scale in this universe; a second one is given by the size of the comoving (homogeneous) Hubble horizon, $d_H/a|_{\epsilon=0}$.  Measuring the latter in units of the former produces a dimensionless ratio,
\begin{equation}
n(t) \equiv {d_H/a|_{\epsilon=0}\over\lambda} = {3\over 2} {k\over\bar a_0} t^{1/3} . 
\label{ratio}
\end{equation}
At times, it is a little more convenient to absorb certain recurring factors, so define $N(t)$ to be a bit more than four times this ratio,
\begin{equation}
N(t) \equiv {4\pi\over 3} {d_H/a|_{\epsilon=0}\over\lambda} 
= {2\pi k\over\bar a_0} t^{1/3} . 
\label{Ntdef}
\end{equation}
Counting the number of wavelengths per horizon thus, the leading behavior for each term of a distinct time-dependence in the metric becomes 
\begin{eqnarray}
a(t,x) &\!\!\!\!=\!\!\!\!& 
\bar a_0 t^{2/3} \Bigl\{
1 + {\textstyle{1\over 20}} N^2 \epsilon^2 \sin^2(2\pi kx) 
\nonumber \\
&& 
- {\textstyle{3\over 2800}} N^4 \epsilon^4 \sin^4(2\pi kx)  
+ \cdots \Bigr\}
\nonumber \\
b(t,x) &\!\!\!\!=\!\!\!\!& 
\bar a_0 t^{2/3} \Bigl\{
1 - {\textstyle{3\over 10}} N^2 \epsilon \cos(2\pi kx) 
\nonumber \\
&& 
+ {\textstyle{9\over 700}} N^4 \epsilon^3 \cos(2\pi kx) \sin^2(2\pi kx) 
+ \cdots \Bigr\} . \quad
\label{abasN}
\end{eqnarray}
The departure from approximate homogeneity occurs first in the $x$-direction, as should have been anticipated, when the number of wavelengths per horizon reaches $N\sim \epsilon^{-1/2}$.

\section{A possible misattribution}
\label{misattribution}

The question now arises---what mistake is made in assuming that the growth of the inhomogeneities has no effect on the large-scale expansion of the universe?  While the inhomogeneous terms evidently grow much faster than the homogeneous part, it might be thought that on average their effect vanishes---if the extra density in one place tends to slow the expansion a little, the surrounding voids will hasten it.  At linear order in the amplitude $\epsilon$, since all of the spatial dependence is a cosine, the spatial average of this function does vanish; however, this property is not true for the higher order terms.  Even after averaging over the spatial coordinates, the effects of the inhomogeneities will remain.  If one then insists on seeing the large-scale expansion as that produced by uniformly distributed matter, it will appear that the expansion has unaccountably begun to speed up. 

The periodicity of this fluctuation means that it is quite simple to define spatially averaged quantities.  For a general function $f(t,x)$, define $\bar f(t)$ to be the result of averaging it over a wavelength,
\begin{equation}
\bar f(t) = \langle f(t,x) \rangle \equiv k \int_0^{1/k} dx\, f(t,x) . 
\label{average}
\end{equation}
Since the metric has a spatial dependence on only one of the coordinates, it will remain anisotropic even after averaging over the $x$-dependence.  Defining therefore two scaling functions $\bar a(t)$ and $\bar b(t)$ for the averaged metric, $\bar g_{\mu\nu}(t)=\langle g_{\mu\nu}(t,x) \rangle$, 
\begin{equation}
ds^2 = dt^2 - \bar b^2(t)\, dx^2 - \bar a^2(t)\, \bigl[ dy^2 + dz^2 \bigr] . 
\label{averagemetric}
\end{equation}
they are found to be given by 
\begin{widetext}
\begin{eqnarray}
\bar a(t) &\!\!\!\!\equiv\!\!\!\!& 
\langle a^2(t,x) \rangle^{1/2} = 
\bar a_0 t^{2/3} \Bigl\{ 
1 + \epsilon^2 \Bigl[ {\textstyle{1\over 40}} N^2 
+ {\textstyle{1\over 36}} \Bigr]
- \epsilon^4 \Bigl[ {\textstyle{11\over 44800}} N^4 
- {\textstyle{1\over 180}} N^2 + {\textstyle{1\over 2592}} \Bigr]
+ {\cal O}(\epsilon^5) 
\Bigr\}
\label{averaged} \\
\bar b(t) &\!\!\!\!\equiv\!\!\!\!& 
\langle b^2(t,x) \rangle^{1/2} = 
\bar a_0 t^{2/3} \Bigl\{ 
1 + \epsilon^2 \Bigl[ {\textstyle{9\over 400}} N^4 
- {\textstyle{1\over 40}} N^2 - {\textstyle{5\over 36}} 
\Bigr] 
- \epsilon^4 \Bigl[ {\textstyle{81\over 320000}} N^8 
- {\textstyle{9\over 112000}} N^6 - {\textstyle{909\over 44800}} N^4 
+ {\textstyle{17\over 360}} N^2 - {\textstyle{5\over 2592}} \Bigr]
+ {\cal O}(\epsilon^5) 
\Bigr\} . 
\nonumber 
\end{eqnarray}
\end{widetext}
Among the terms with the same time-dependence, which is implicit in the $N(t)$ dependence, the leading terms at a particular order in $N$ and $\epsilon$ are 
\begin{eqnarray}
\bar a(t) &\!\!\!\!=\!\!\!\!& 
\bar a_0 t^{2/3} \Bigl\{ 
1 + {\textstyle{1\over 40}} N^2 \epsilon^2 
- {\textstyle{11\over 44800}} N^4 \epsilon^4 
+ \cdots  
\Bigr\}
\nonumber \\
\bar b(t) &\!\!\!\!=\!\!\!\!& 
\bar a_0 t^{2/3} \Bigl\{ 
1 + {\textstyle{9\over 400}} N^4 \epsilon^2
- {\textstyle{81\over 320000}} N^8 \epsilon^4
+ \cdots 
\Bigr\} . \quad 
\label{leadaveraged} 
\end{eqnarray}

Although $\bar a(t)$ has no true analogue in the actual universe, since there is no genuine translation invariance in any direction, even its behavior is seen to stray from a $t^{2/3}$ expansion when $N$ grows to be the same size as $\epsilon^{-1}$.  The importance of the inhomogeneities appears much sooner in the $x$ direction, when $N\sim\epsilon^{-1/2}$.  When comparing to physical scales in the universe, it is important to remember that $N(t)$ is not quite equal to the number of wavelengths per horizon, which was instead called $n(t)$, nor is $\epsilon$ equal to the initial amplitude in the physical density of the matter.  If at the time $t_0$, the wavelength of the fluctuation is exactly equal to the Hubble horizon ($n(t_0)=1$), then the amplitude of the initial fluctuation in the physical density, 
\begin{equation}
\rho(t_0,x) = {1\over 6\pi G} {1\over t_0^2} \bigl[ 1 + \delta\cos(2\pi kx) + {\cal O}(\epsilon^2) \bigr] , 
\label{deltadef}
\end{equation}
which has been called $\delta$ to distinguish it from the amplitude in the comoving density, is a bit larger than $\epsilon$, 
\begin{equation}
\delta = {8\pi^2\over 15} \epsilon .
\label{deltaaseps}
\end{equation}
Replacing $N$ and $\epsilon$ with $n$ and $\delta$ does not much alter the form of the scaling factor, 
\begin{equation}
\bar b(t) = \bar a_0 t^{2/3} 
\Bigl\{ 1 + {\textstyle{1\over 4}} n^4 \delta^2 
- {\textstyle{1\over 32}} n^8 \delta^4 + \cdots \Bigr\} .
\label{btindelta}
\end{equation}
Once $n^2\delta$ has grown larger than about ${}^{\hbox{\scriptsize 2}}\!\!\slash\!${\scriptsize 3}, the expansion rate begins to increase.

Remember that the horizon size tends to grow at a faster rate than the physical wavelength of the fluctuation, at least during the early stage when the expansion rate is governed by the homogeneous term.  Beyond this stage, while $n(t)$ can still be formally defined, it corresponds less and less to the meaning previously assigned to it.  But since this departure sets in fairly gradually, and since from a cosmological perspective it began comparatively recently, $n$ should still very roughly correspond to the number of wavelengths per horizon.

The threshold between the era governed by the homogeneous term and that when inhomogeneities become more important for the overall expansion occurs when $n^2\delta\sim 1$.  Just prior to this threshold, the expansion rate begins to accelerate as it passes from a $t^{2/3}$ dependence to a $t^{2/3}n^4\sim t^2$ dependence.  Beyond this point, the solution here is not sufficient for determining the extent of this acceleration and whether and how long it lasts, since a perturbative description of the metric based solely on the smallness of $\delta$ or $\epsilon$ fails.

To see, from a slightly different perspective, how the departure from homogeneity leads to an acceleration in the expansion rate of the universe, consider the first of the components of Einstein's equation, which in a homogeneous universe corresponds to Friedmann's equation, 
\begin{equation}
G_{00} \equiv R_{00} - {\textstyle{1\over 2}} g_{00} R = 8\pi G\, T_{00} . 
\label{einstein00}
\end{equation}
The spatial average of the left side leads to a more complicated apparent expansion than the $t^{-2}$ scaling that could be explained by a uniformly distributed gas of matter alone, 
\begin{eqnarray}
\bar G_{00} &\!\!\!\!=\!\!\!\!&
{\textstyle{4\over 3}} t^{-2} \Bigl\{ 
1 + {\textstyle{1\over 8}} \epsilon^2 
\Bigl[ {\textstyle{9\over 25}} N^4 - N^2 \Bigr]
\nonumber \\
&&\qquad
+ {\textstyle{1\over 40}} \epsilon^4 \Bigl[ {\textstyle{243\over 2000}} N^8 
- {\textstyle{3069\over 2800}} N^6 + {\textstyle{2209\over 560}} N^4 
- {\textstyle{41\over 9}} N^2 \Bigr]
\nonumber \\
&&\qquad
+ {\cal O}(\epsilon^5) 
\Bigr\} . 
\label{averaged00}
\end{eqnarray}
Most of these terms are only tiny corrections, or become important only far into the regime when the inhomogeneities have grown large.  Immediately before the transition into this regime, the leading part of the gravitational part of the averaged Friedmann's equation is 
\begin{equation}
\bar G_{00}(t) =
{4\over 3} {1\over t^2} \biggl\{ 
1 + {1\over 2} n^4 \delta^2
+ {3\over 8} n^8 \delta^4
+ \cdots  
\biggr\} . 
\end{equation}
If one insists on accounting for this behavior only through uniformly distributed materials, ordinary matter alone would appear not to be sufficient.  By itself, a uniform gas of matter would explain the first term, but other materials would need to be invoked to account for the other terms, both of which are diluted more slowly than the first.  The second term scales as $n^4/t^2\sim t^{-2/3}$, while if some strange substance were introduced to explain the third, its density would appear to be growing with the expansion, as $n^8/t^2\sim t^{2/3}$.  However, this would be a little misleading; once the universe has passed the threshold were $n^2\delta\sim 1$, the actual behavior of this artificial substance would be determined by adding together all of the terms scaling as arbitrary powers of $n^2\delta$, at least until the next threshold when the terms scaling as powers of $n^3\delta^2$ become important too.

In this example no such material is needed.  It would only be invoked by insisting that the large-scale expansion is governed by a uniformly distributed substance and thereby misattributing the origin of the change in the expansion rate.  The source of the acceleration and the advent of the new expansion rate are due here to the inhomogeneities in the distribution of the matter alone.

The fact that the average of the scale factor accelerates does not by itself immediately mean that the luminosity distance function, which is measured through the observations of distant supernovae, behaves as it does in the actual universe.  The calculation of this function will be left to later work, since it is important first to understand the behavior well into the inhomogeneous era, which probably needs to be studied numerically, to see the extent of the acceleration, both in terms of its overall magnitude and its duration.

\section{Physical scales}
\label{scales}

This example of an inhomogeneous universe was constructed to be a simple as possible, while still keeping the essential features of the actual universe that were to be analyzed.  The results of the calculation showed that once the inhomogeneities have matured sufficiently, they can significantly affect the large-scale expansion, which begins to accelerate as it passes into this later stage.  This effect is found by calculating the higher order, nonlinear behavior of the solution to Einstein's equations; but because of the inherent smallness of the initial amplitude of the fluctuations, a fair amount of time must first pass before this influence appears, during which the horizon grows much larger than the wavelength of the fluctuation.  So the idea that because the resulting structures are small compared with the size of the observable universe they are unimportant is a mistaken one---at least in this example---or rather it is true only while nonlinear parts of the solution are still small.

The expansion of the actual universe seems to have departed from the behavior expected of a space-time uniformly filled with matter about four billion years ago or so.  If this departure was produced by some of the regularly occurring larger structures in the matter, then their size at that time should be simply related to the amplitude of the initial fluctuation, $\delta$.  From what can be inferred from the observations of the Wilkinson Microwave Anisotropy Probe \cite{wmap}, the size of the product of the change in the spatial curvature from its average value, evaluated at two different places, is $(2.430\pm 0.091)\times 10^{-9}$, with only a slight dependence on the distance between the two points.  If the initial amplitude of the matter's density is taken to be comparable to the square root of this value, so that $\delta\sim 5\times 10^{-5}$, then the stage of the expansion that is strongly affected by the inhomogeneities ought to have begun when the structures responsible for it had a typical size so that $n\sim 140$.  To evolve their size from its value four billion years ago to today requires knowing the time-dependence of the metric in the era governed by the inhomogeneities, which is not possible for the approximation used in this calculation.  However, assuming that the expansion rate, which has only changed fairly recently according to a cosmological reckoning, does not depart excessively from the $n\sim t^{1/3}$ dependence of the homogeneous era before it, the size of the structures responsible for the acceleration would now be ${}^{\hbox{\scriptsize 1}}\!\!\slash\!${\scriptsize 160} the size of the horizon.  The appropriate horizon is not the size of the observable universe today, which is estimated to be about 14.5 Gpc, but is rather the Hubble horizon.  This horizon is inversely proportional to the current value of the Hubble parameter, $H_0$, used to compute the rate of the recessional velocity of distant objects.  It is also measured under the assumption of a homogeneous cosmology on average.  The size of this Hubble horizon, $c/H_0$, is approximately 4 Gpc,  which in turn implies that the size of the structures that are chiefly responsible for the change in the recent expansion history of the universe ought to be about 30 Mpc today, which is comparable to the size of the smaller voids.

In the actual universe, the voids are fully three dimensional and not periodic in one dimension only, nor are they arranged in a simple periodic array, nor is there a unique size of the voids.  Their size today typically falls within the range of about 40--90 Mpc \cite{voids}.  When the Hubble horizon is measured in units of this range of void-widths, and then followed back to four billion years ago, still very crudely applying the $t^{1/3}$ dependence, the range for the values of $n$ falls between 40 and 90.  This range corresponds to a range of values for $n^2\delta$, 
\begin{equation}
0.1 \lesssim n^2\delta \lesssim 0.4 , 
\label{n2delta}
\end{equation}
which is consistent with the expectation that the acceleration should have begun once $n^2\delta\sim {}^{\hbox{\scriptsize 2}}\!\!\slash\!${\scriptsize 3}.

Whether or not this mechanism is in fact responsible for the relation that has been observed between the luminosities and the redshifts of distant supernovae requires improving the example presented here so that it includes a few more of the properties of the actual universe.  Enlarging the spatial dependence to include all of the coordinates and allowing for a less regular array of voids probably will not alter much the conclusions from this simpler periodic example.  It is probably more important to include a full spectrum of fluctuations.  Shorter wavelengths have been growing for a longer time than the scales associated with the voids, so it might be expected that the effects of the inhomogeneities at smaller scales ought to have appeared earlier.  In its full, three-dimensional momentum representation, the power spectrum for the primordial fluctuations inferred from observations decays approximately as $k^{-3}$.  A smaller amplitude for shorter wavelengths would provide a natural suppression in $\delta$ that would compensate for the longer time, $n$, that shorter wavelengths would have been smaller than the Hubble horizon.  This effect would additionally reduce the role of mixings among different wavelengths, which would also need to be evaluated when the metric is treated well beyond a linear approximation.  Once these properties have each been addressed, it should then be possible to evaluate how the true inhomogeneities affect what is actually observed.  

The encouraging conclusion of this work is that it shows that it might not be necessary to abandon altogether the idea that the universe is composed primarily of matter nor that its dynamics are governed by ordinary general relativity, in which case the discrepancies between some of the earlier theoretical expectations and observations would rather be due to having insisted on assumptions that do not hold in the actual universe.

\vskip12truept

The author would like to express his thanks for the support provided by the Niels Bohr International Academy and the Discovery Center at the Niels Bohr Institute.  This calculation was motivated by a talk by Sysky R\"as\"anen at the {\small COSMO}-09 conference, which was based on some of his earlier work \cite{cosmo09}.

\end{document}